\documentclass{PoS}

\usepackage{array}
\usepackage{epsfig}
\usepackage{amssymb}
\usepackage{graphics,graphpap}
\usepackage{amssymb}
\usepackage{amsmath}

\title{New physics in $\Delta \Gamma_d$}

\ShortTitle{New physics in $\Delta \Gamma_d$}

\author{\speaker{Gilberto Tetlalmatzi-Xolocotzi}\\
        Institute for Particle Physics and Phenomenology, Durham University\\
        E-mail: \email{gilberto.tetlalmatzi-xolocotz@durham.ac.uk}}

\abstract{We analyze the possibility of having new physics effects in the decay rate difference, $\Delta \Gamma_d$,
          of neutral $B_d$ mesons.  Three different sources of enhancement are considered, CKM unitarity violations,
          beyond standard model effects in the tree-level dimension six operators $(\bar{d}p)(\bar{p}'b)$ with $p,p'=u,c$; 
          and large enhancements of the almost unconstrained operators $(\bar{d}b)(\bar{\tau}\tau)$. 
          We find that deviations of several hundred per cent from the standard model prediction of $\Delta \Gamma_d$
          are not excluded by current experimental data.
           }

\FullConference{Flavorful Ways to New Physics - FWNP,\\
		28-31 October 2014\\
		Freudenstadt - Lauterbad, Germany}

\begin{document}
\section{Introduction}

There is a longstanding discrepancy between experimental results for the like-sign dimuon asymmetry measured by the D0 Collaboration
\cite{D01}-\cite{D0_exp} and the corresponding standard model predictions \cite{theo_Lenz_Uli}-\cite{theo_4}.
In \cite{Borissov} an interesting connection between the dimuon asymmetry and the decay rate difference of neutral $B_d$ mesons was
suggested: the measured enhancement of the dimuon asymmetry could also be explained by an enhanced decay rate difference of the neutral
$B_d$ mesons. Moreover, $\Delta  \Gamma_d$, is currently only weakly constrained by direct measurements.
This was the motivation for the study in \cite{Delta_Gamma_d} where indirect experimental constraints on possible new physics enhancements
of $\Delta \Gamma_d$ were studied; we present here the main findings of \cite{Delta_Gamma_d}.

\section{Neutral $B$ mixing}

Due to electroweak interactions the neutral states $B_d$ and $\bar{B}_d$ oscillate into each other, the time evolution of this
system is given by solving the Schr\"{o}dinger-like equation 
\vspace{-0.6cm}
\begin{center}\begin{eqnarray}
i\frac{d}{dt}\begin{pmatrix}\left|B_d\right>  \\ \left|\bar{B_d}\right>\end{pmatrix} = 
\Sigma^d \begin{pmatrix}\left|B_d\right> \\  \left|\bar{B_d}\right>\end{pmatrix}\hspace{0.4 cm} \hbox{  where    }\hspace{0.4 cm} \Sigma^d = M_d-\frac{i}{2}
\Gamma_d=\begin{pmatrix} M^d_{11}-\frac{i\Gamma^d_{11}}{2} & M^d_{12}-\frac{i\Gamma^d_{12}}{2}\\ 
M_{12}^{d*}-\frac{i\Gamma_{12}^{d*}}{2} & M^d_{11}-\frac{i\Gamma^d_{11}}{2} \end{pmatrix}.
\end{eqnarray}
\end{center}
\noindent
Diagonalizing $\Sigma^d$ gives the physical eigenstates $\{|B_H>, |B_L>\}$ with the masses $M^d_H, M^d_L$ and the decay rates
$\Gamma^d_H, \Gamma^d_L$.
\noindent
To provide a mathematical description of the mixing process it is useful to define the observables $\Delta M_d = M^d_H-M^d_L$
and $\Delta \Gamma_d =  \Gamma^d_H-\Gamma^d_L$. Theoretically $\Delta M_d$ and $\Delta \Gamma_d$ can be calculated
 from the components of $\Sigma^d$ according to the formulas 

\vspace{-0.6cm}

\begin{center}
\begin{eqnarray}
\label{eq:DeltaMGamma}
\Delta M_d \approx 2|M^d_{12}|\hspace{0.4 cm} \hbox{  and  }\hspace{0.4 cm} \Delta \Gamma_d \approx 2|\Gamma^d_{12}|cos(\phi_d)
\hspace{0.4 cm} \hbox{  where  }\hspace{0.4 cm} \phi_d=arg\Bigl(-\frac{M^d_{12}}{\Gamma^d_{12}}\Bigl).
\end{eqnarray}
\end{center}
\noindent
Only $\Delta M_d$ and $\Delta \Gamma_d$ are directly accessible in experiment, the phase $\phi_d$ can be calculated 
from the measurement of the semileptonic asymmetry
\vspace{-0.6cm}
\begin{center}
\begin{eqnarray}
\label{eq:asymm}
a^d_{sl}=\Bigl|\frac{\Gamma^d_{12}}{M^d_{12}}\Bigl| sin(\phi_d).
\end{eqnarray}
\end{center}

\noindent
By combining data from Belle, BABAR, D0, DELPHI and LHCb the following direct experimental bound for $\Delta \Gamma_d$ is available 

\begin{eqnarray}
\frac{\Delta \Gamma^{HFAG}_d}{\Gamma_d}&=&(0.1 \pm 1.0) \%\cite{HFAG}
\end{eqnarray}


\noindent
which can be compared to the standard model prediction

\begin{eqnarray}
\frac{\Delta \Gamma_d^{SM}}{\Gamma_d}&=&(0.42\pm 0.08)\%\cite{theo_Lenz_Uli}.\nonumber
\end{eqnarray}


\section{Sources of enhancement for $\Delta \Gamma_d$}
\subsection{CKM unitarity violations}

Let $\lambda^d_q=V^*_{qd}V_{qb}$, from the unitarity of the $3\times3$  CKM matrix the following condition is satisfied 
$\lambda^d_u+\lambda^d_c+\lambda^d_t=0$. However in different extensions of the standard model the previous equality
is broken according to $\lambda^d_u+\lambda^d_c+\lambda^d_t=\delta^d_{CKM}$; for instance in 4th family studies  
$\delta^d_{CKM}\approx 10^{-2}$ leading to a potential enhancement factor on $\Delta \Gamma_d$ of $300\%$.

\subsection{Current-current standard model operators}

In this section we consider the possibility of having new physics effects in tree level standard model operators and its
consistency with recent experimental observations, a similar study and its implications over the CKM phase $\gamma$ has been
discussed in \cite{gamma}.\\
\\
\noindent
We are concerned with the following effective Hamiltonian

\begin{eqnarray}
\label{eq:EffH}
\mathcal{H}_{eff}^{current, |\Delta B|=1}&=&\frac{4G_F}{\sqrt{2}}\sum_{p,p'=u,c}V^{*}_{pd}V_{p'b}\sum_{i=1,2}C_{i}^{pp'}(\mu)Q_{i}^{pp'}+h.c.
\end{eqnarray}
\noindent
with

\begin{figure}[!t]
\begin{center}
\vspace{-5mm}
  \includegraphics[height=0.425 \textwidth]{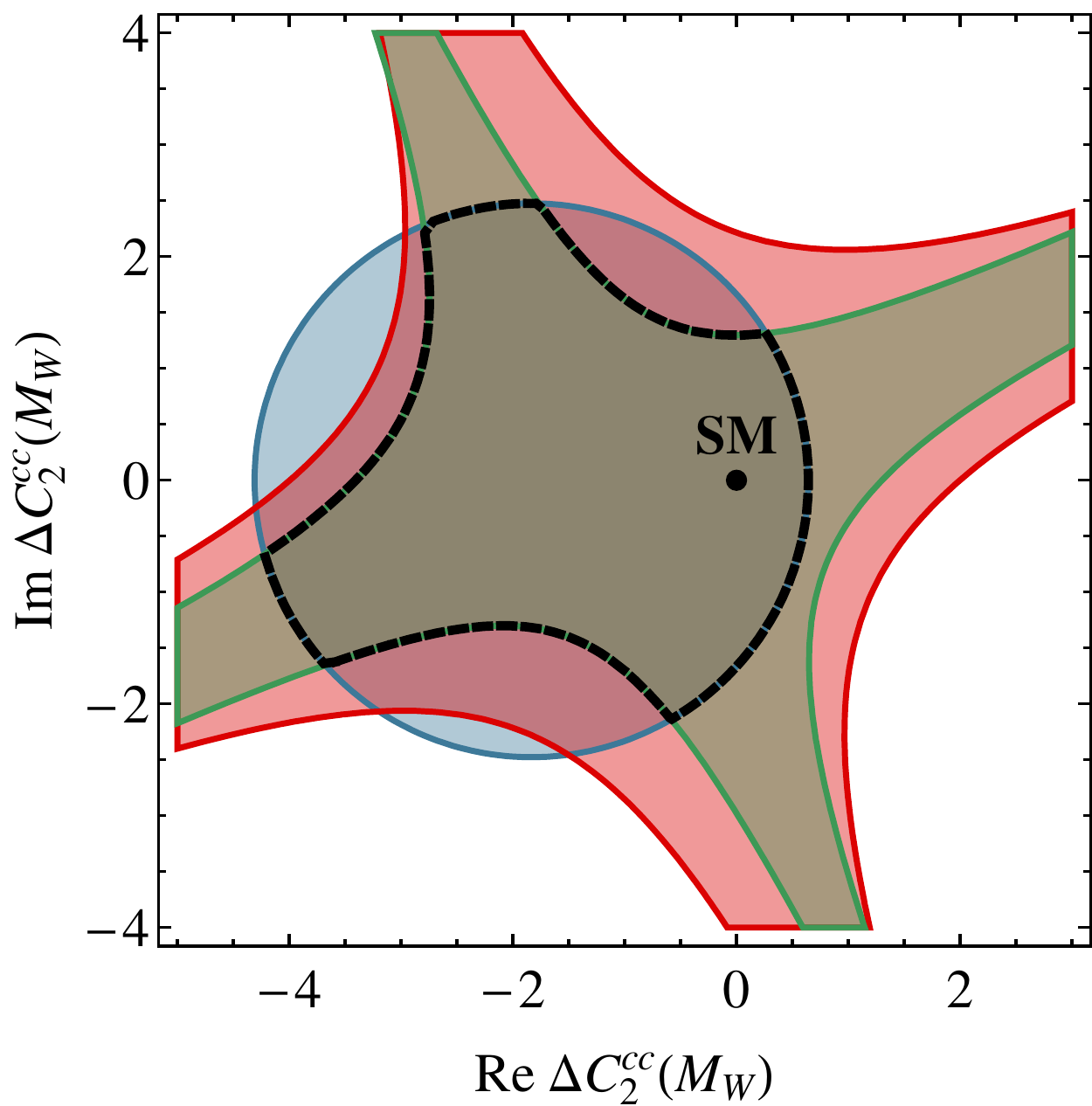} \qquad
  \includegraphics[height=0.425 \textwidth]{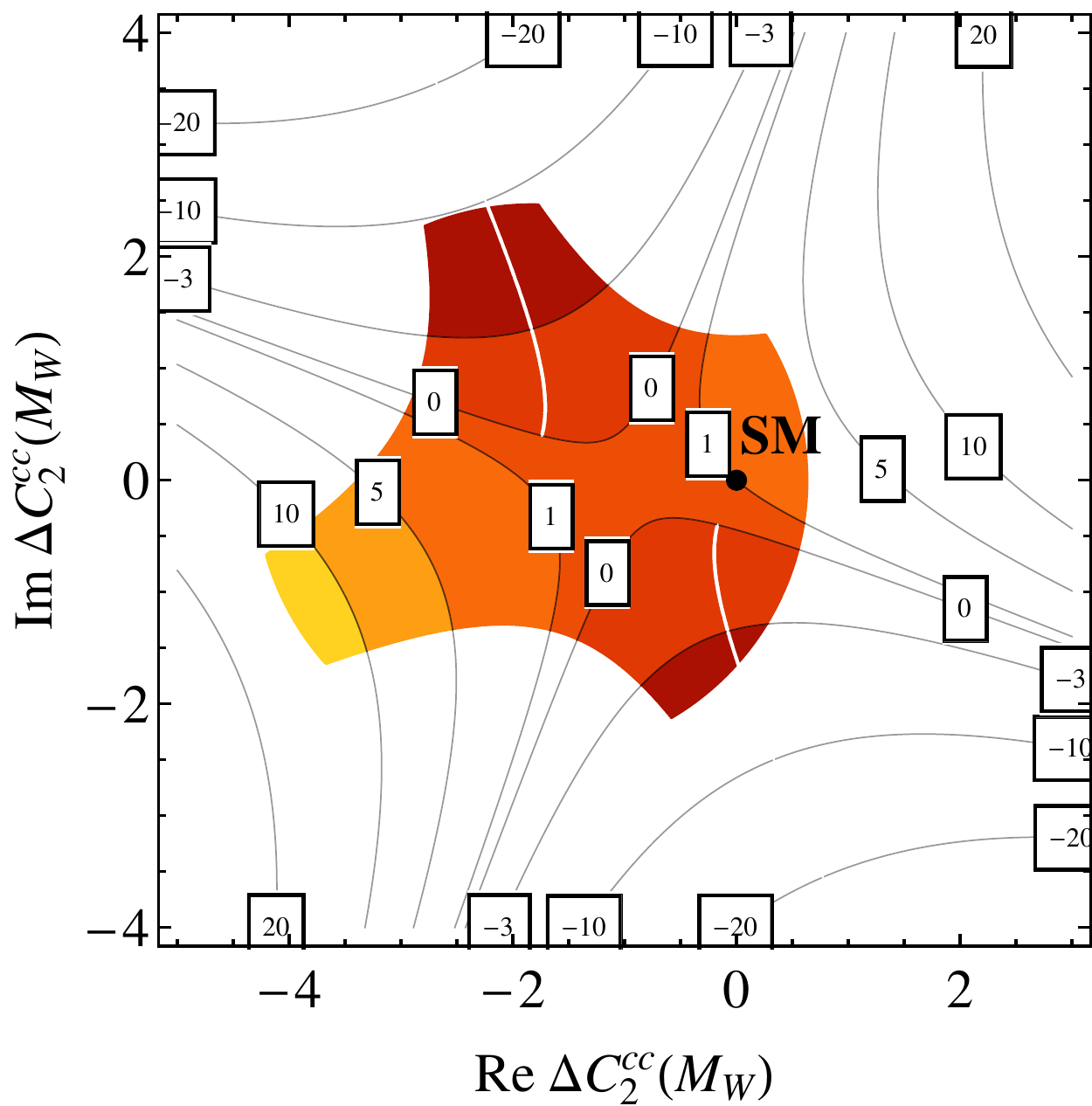} \\[4mm]
\end{center}
\caption{Left panel: Allowed parameter space in the Re $\Delta C_2^{cc}$-Im$\Delta C^{cc}_{2}$ plane. The constraints shown
correspond to $B\rightarrow X_d \gamma$ (Blue), $a^d_{sl}$ (Green) and $sin(2\beta)$ (Red). The region contained within the dashed 
black line represent the combined constraints from the different observables.  Right panel: contours of $\Delta \Gamma_d/\Delta \Gamma^{SM}_d$
along with the combined constraints.}
\label{fig:C2}
\end{figure}

\vspace{-0.6cm}

\begin{eqnarray}
\label{eq:tree}
 Q_1^{pp'} &=& (\bar{d}^\alpha p^\beta)_{V-A} (\bar{p'}^{\beta}b^{\alpha})_{V-A} \; ,\nonumber
\\
 Q_1^{pp'} &=& (\bar{d}^\alpha p^\alpha)_{V-A} (\bar{p'}^{\beta}b^{\beta})_{V-A}   \;.
\end{eqnarray}
\noindent
New physics effects will produce shifts over the standard model Wilson coefficients according to 

\begin{eqnarray}
\label{eq:shiftsC}
C^{pp'}_{1,2}&=&C_{1,2}^{SM}+\Delta C^{pp'}_{1,2}.
\end{eqnarray}
\noindent
To constrain the values of $\Delta C^{pp'}_{1,2}$ we consider different observables depending on the up type quark structure
indicated by the labels $pp'$ in Eqns.  (\ref{eq:EffH})  and (\ref{eq:tree}).
The most important bounds arise from the operators related with the transitions
$b\rightarrow u\bar{u}d$, $b\rightarrow c\bar{u}d$ and $b\rightarrow c\bar{c}d$ with Wilson coefficients $C^{uu}$, $C^{uc}$ and
$C^{cc}$ respectively.
\noindent
The decays $B\rightarrow \pi\pi$, $B\rightarrow \pi\rho$ and $B\rightarrow \rho\rho$ provide limits for  $\Delta C^{uu}$.
To calculate the bounds on $\Delta C^{uc}$ the relevant processes are $\bar{B}\rightarrow D^{*+}\pi^-$
and $B^0\rightarrow D^{(*)0} h^0$ with $h^0=\pi^0, \eta \hbox{ or } \omega$. Regarding that the decay width of the $B_d$ meson is dominated by the transition $b\rightarrow c\bar{u}d$ an extra constraint for $\Delta C^{uc}$ can be imposed from the decay rate itself.
\noindent
The shift $\Delta C^{cc}$ is bounded by the process $B\rightarrow X_d\gamma$, the observable $sin(2\beta)$
and the semileptonic asymmetry $a_{sl}^d$ defined in Eqn. (\ref{eq:asymm}). We visualize graphically the parameter space available
 for new physics by plotting the real and imaginary parts of $\Delta C_{1,2}^{qq'}$, as an example we present the results for $\Delta C_2^{cc}$
in the left panel of Fig. (\ref{fig:C2}).

\noindent
\\
Finally the effects of $\Delta C_{1,2}^{qq'}$  over $\Delta \Gamma_d$ can be estimated by using  Eqns. (\ref{eq:DeltaMGamma}) and
(\ref{eq:shiftsC}) together with the expression

\begin{eqnarray}
\label{eq:Gamma12}
\Gamma_{12}^d&=&\frac{1}{2M_{B_d}}<\bar{B_d}| Im \left( i\int d^4x \hat{T}\left[\mathcal{H}_{eff}^{current, \Delta B=1}(x)\mathcal{H}_{eff}^{current, \Delta B=1}(0)\right]\right)|B_d>\nonumber\\
&=&- \left(  \lambda_c^2     \Gamma_{12}^{cc,d}(C_1^{cc},C_2^{cc}) +
                2 \lambda_c \lambda_u \Gamma_{12}^{uc,d}(C_1^{uc},C_2^{uc}) +
                  \lambda_u^2     \Gamma_{12}^{uu,d}(C_1^{uu},C_2^{uu}) \right). \;
\end{eqnarray}
\\
\noindent
The enhancement factors can be read by plotting the contour lines for $\Delta \Gamma_d/\Delta \Gamma^{SM}_d$,  we provide an example 
of this strategy in the right panel of Fig. (\ref{fig:C2}).
\noindent
Our calculations show that there is room for new physics in $C_{1,2}^{uu}$ and $C_{1,2}^{uc}$  leading to enhancement factors of $1.6$ 
in both cases. The most interesting result arises from  $C_{2}^{cc}$ where 
an enhancement factor of 14 is allowed by current experimental results. All the bounds were calculated at $90\%$ C.L..

\subsection{Operators $(\bar{d}b)(\bar{\tau} \tau)$}

In the third part of our analysis we consider the effective operators $(\bar{d}b)(\bar{\tau} \tau)$, they are well suppressed within the standard model,  however they are not quite constrained by the  experimental data available nowadays; the approach followed in this section is analogous to the study performed for $B_s$ mesons in \cite{Bobeth_Haisch}.
\noindent
Here we take into account the following effective Hamiltonian

\begin{eqnarray}
\mathcal{H}^{b\rightarrow d\tau^+\tau^-}_{\rm eff}= -\frac{4G_F}{\sqrt{2}}\lambda_t^d\sum_{i,j} C_{i,j}(\mu)Q_{i,j}
\end{eqnarray}

\noindent
and study the scalar, vector and tensor Dirac structures of $(\bar{d}b)(\bar{\tau}\tau)$

\begin{eqnarray}
  \label{eq:operators}
  Q_{S, AB} &=& \left (\bar d \, P_A \, b \right ) \left (\bar \tau
    \, P_B \, \tau \right ) \,,\nonumber\\
  Q_{V, AB} &=& \left (\bar d \, \gamma^\mu P_A \, b \right )
  \left (\bar \tau\, \gamma_\mu P_B \, \tau \right ) \,, \\
  Q_{T, A} &=& \left (\bar d \, \sigma^{\mu\nu} P_A \, b \right )
  \left (\bar \tau \, \sigma_{\mu \nu} P_A \,\tau \right ) \nonumber
\end{eqnarray}

\noindent
with Wilson coefficients $C_{S,AB}$, $C_{V,AB}$ and $C_{T,A}$ respectively. Following the notation of \cite{Bobeth_Haisch}: $A,B=R,L$ and $P_{R,L}=(1\pm \gamma_5\Bigl)/2$.\\
\\
\noindent
To get bounds on the Wilson Coefficients we use direct and indirect constraints. Direct bounds arise from particle transitions
where the chain $b\rightarrow d \tau^+\tau^-$ appears at tree level; indirect estimations are derived from processes where the terms
given in Eqns. (\ref{eq:operators}) arise as a consequence of operator mixing, loop level corrections or both. 
\noindent
The direct category includes the processes $B_d\rightarrow \tau^+ \tau^-$,  $B\rightarrow X_d \tau^+\tau^-$ and
 $B^+ \rightarrow \pi^+ \tau^+\tau^-$. In the first case the following experimental bound is available 
 $Br(B_d\rightarrow \tau^+ \tau^-)<4.1\times 10^{-3}$ \cite{BABAR_bound}. In the second and third cases currently there are not experimental 
measurements for the branching ratios. However we can estimate an upper limit for these observables by considering the space available
for new $B_d$ decays, i.e. those that have not been measured yet. To fulfill this task we compare the ratio $\Bigl(\frac{\tau_{B_s}}{\tau_{B_d}}-1\Bigl)_{SM}$ against $\Bigl(\frac{\tau_{B_s}}{\tau_{B_d}}-1\Bigl)_{exp}$ and ignore new physics effects in the $B_s$ sector; our result is  
$Br(B_d\rightarrow X_{new})\leq 1.5 \%$. 
\noindent
From $B_d\rightarrow \tau^+ \tau^-$ we get chirality independent bounds for $C_S$
and $C_V$, whereas from $B\rightarrow X_d \tau^+\tau^-$ and  $B^+ \rightarrow \pi^+ \tau^+\tau^-$ we get constraints for
all the Dirac structures shown in Eqns. (\ref{eq:operators}).
\noindent
Indirect bounds are calculated from the branching ratios of the processes $B_d\rightarrow X_d \gamma$ and $B^+\rightarrow \pi^+\mu^+\mu^-$. 
In the case of $B_d\rightarrow X_d \gamma$ the tensor element in Eqns. (\ref{eq:operators}) mixes with the operator mediating the transition
$b\rightarrow d \gamma$. For the process $B^+\rightarrow \pi^+\mu^+\mu^-$ there are contributions from mixing between the vector
and tensor components of Eqns. (\ref{eq:operators}) and the operators responsible for the chains $b\rightarrow d \gamma$ and
$b\rightarrow d l^+ l^-$.\\
\\
\noindent
To quantify the effects of our $(\bar{d}b)(\bar{\tau} \tau)$ operators over $\Delta \Gamma_d$ we first parametrize the new physics
contributions in $\Gamma^d_{12}$ through the factor $\tilde{\Delta}_d$

\vspace{-5mm}

\begin{eqnarray}
\Gamma_{12}^d&=&\Gamma_{12}^{d,SM}\tilde{\Delta}_d.
\end{eqnarray}

\noindent
The relationship between $\tilde{\Delta_d}$ and the Wilson coefficients of the operators in Eqns. (\ref{eq:operators})
is established by the following inequalities

\vspace{-5mm}
\begin{eqnarray}
\label{eq:optautau}
|\tilde{\Delta}_d|_{S,AB}&<&1+(0.41^{+0.13}_{-0.08})|C_{S,AB}(m_b)|^2\nonumber\\
|\tilde{\Delta}_d|_{V,AB}&<&1+(0.42^{+0.13}_{-0.08})|C_{V,AB}(m_b)|^2\\
|\tilde{\Delta}_d|_{T,A}&<&1+(0.42^{+0.13}_{-0.08})|C_{T,A}(m_b)|^2\nonumber
\end{eqnarray}

\noindent
here we are assuming single operator dominance; i.e. we are analyzing the effect of only one operator from Eqns. 
(\ref{eq:operators}) at a time by switching on the corresponding Wilson coefficient and setting the others to zero. The subindex
of $|\tilde{\Delta}_d|$ in Eqns. (\ref{eq:optautau}) indicates the operator under study.\\
\\
\noindent
We finally arrive to the following enhancement factors at $90\%$ $C.L.$

\begin{figure}[!t]
\begin{center}
\vspace{-5mm}
  \includegraphics[height=0.425 \textwidth]{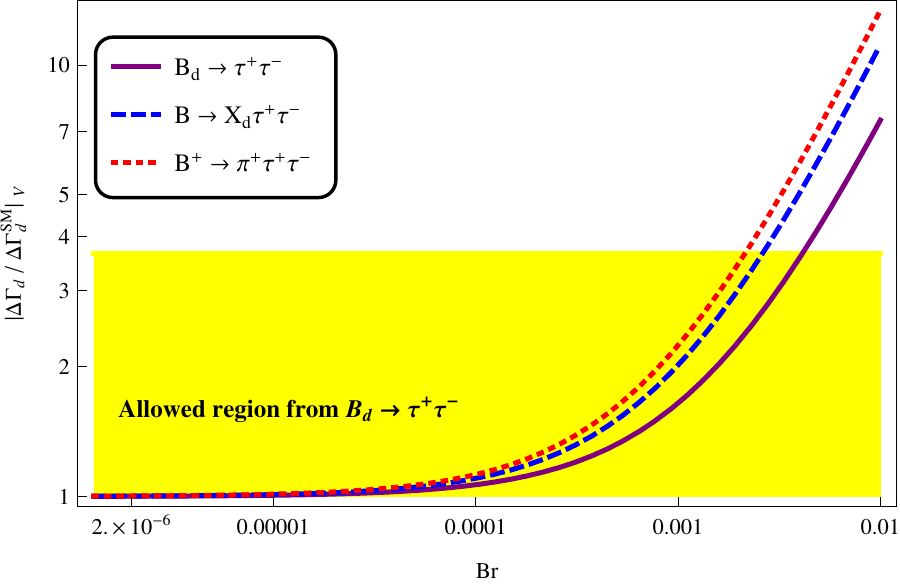} \qquad
\end{center}
\caption{Bounds on the enhancement factor for $\Delta \Gamma_d$ as a function of the improvement on the experimental limits
over the branching ratios for the processes $B_d\rightarrow \tau^+\tau^-$ (purple), $B_d\rightarrow X_d\tau^+\tau^-$ (blue)
and $B_d\rightarrow \pi^+\tau^+\tau^-$ (red). Only the effect induced by the vectorial version of the  $(\bar{d}b)(\bar{\tau}\tau)$
 operators is shown at $90\%$ C. L..}
\label{fig:tautauV}
\end{figure}

\vspace{-5mm}
\begin{eqnarray}
\label{eq:boundstautau}
|\tilde{\Delta}_d|_{S,AB}&\leq& 1.6\nonumber\\
|\tilde{\Delta}_d|_{V,AB}&\leq& 3.7\\
|\tilde{\Delta}_d|_{T,R}&\leq& 1.2.\nonumber
\end{eqnarray}
\noindent
Stronger bounds for $|\tilde{\Delta}_d|$ can be established if upper limits for the branching ratios of the processes  
$B_d\rightarrow \tau^+ \tau^-$,  $B\rightarrow X_d \tau^+\tau^-$ and $B^+ \rightarrow \pi^+ \tau^+\tau^-$ are reduced experimentally;
this effect is shown explicitly in Fig. (\ref{fig:tautauV})  for the vector structure of the $(\bar{d}b)(\bar{\tau}\tau)$ operators.

\section{Like sign dimuon asymmetry}

Experimentally the  D0 collaboration has  measured the raw like sign  dimuon charge
asymmetry $A=A_{CP}+A_{bkg}$ \cite{D01}-\cite{D0_exp}, after removing the backgrounds  $A_{bkg}$ the remaining component $A_{CP}$ is the result of CP violation arising from neutral B meson interactions. Before 2013 it was assumed that $A_{CP}$ was caused only by CP violation 
in mixing between the $B_d$ and $\bar{B}_d$ states leading to

\begin{eqnarray}
\label{eq:ACP1}
A_{CP}\propto A^d_{sl}=C_d a^d_{sl}+C_s a_{sl}^s.
\end{eqnarray}
\noindent
The terms  $a_{sl}^{d,s}$ are  the semileptonic asymmetries and $C_{d,s}$ are proportionality
constants depending on the production fractions, oscillation parameters and decay widths of $B_{d,s}$. Based on this interpretation
D0 reported in 2011 a measurement for $A_{sl}^d$ that was in disagreement with the standard model prediction by $3.9 \sigma$ \cite{D03}.\\
\\
\noindent
Recently the contribution of CP violation in interference between decays with and without mixing was included in the analysis of  $A_{CP}$.
According to \cite{Borissov} this implies the replacement of Eqn. (\ref{eq:ACP1}) by
 
\begin{eqnarray}
\label{eq:ACP2}
A_{CP}\propto A^d_{sl}+C_{\Gamma_d}\frac{\Delta \Gamma_d}{\Gamma_d}
+C_{\Gamma_s}\frac{\Delta \Gamma_s}{\Gamma_s},
\end{eqnarray}
\noindent
here  $\frac{\Delta \Gamma_s}{\Gamma_s}$ is highly suppressed by the constant $C_{\Gamma_s}$
in comparison with the contribution due to  $\frac{\Delta \Gamma_d}{\Gamma_d}$.\\
\\
\noindent
In principle Eqn. (\ref{eq:ACP2}) may suggest that the tension between
theory and experiment for $A_{CP}$ could be eliminated if $\Delta \Gamma_d$ gets enhanced through non standard model physics.
However a more detailed study reveals that Eqn. (\ref{eq:ACP2})  is the first approximation towards the inclusion
of CP violation in interference; instead of having 
a complete dependence on $\Delta \Gamma_d$ it is expected that the relevant contribution to the interference
depends on the components $\Gamma^{cc}_{12}$ and $\Gamma^{uc}_{12}$, see Eqn. (\ref{eq:Gamma12}),
in a more convoluted way than the one given in $\Delta \Gamma_d$ \cite{Uli_dimuon};
this possibility is currently under further investigation \cite{dimuon}.  
 
\section{Conclusions}

We have explored the possibility of having new physics effects on $\Delta \Gamma_d$ within the Heavy Quark Expansion.
Firstly we found that by breaking the unitarity of the CKM matrix by $10^{-2}$ we can have a deviation of $300 \%$
with respect to the standard model expectation on $\Delta \Gamma_d$. Our analysis of dimension six tree level standard model effective operators
shows interesting deviations over $\Delta \Gamma^{SM}_d$,  the most remarkable example is given by $(\bar{d}c)(\bar{c}b)$
where an enhancement factor of 14 is allowed by current experimental data. 
Finally in the case of the effective operators $(\bar{d}b)(\bar{\tau}\tau)$ we have found that $\Delta \Gamma_d$
can be nearly $4$ times bigger than in the standard model. 
We mentioned that as a first approximation there is a relationship between $\Delta \Gamma_d$ and the dimuon asymmetry $A_{CP}$
measured by D0, describing briefly possible corrections that could be useful in reducing the gap between theory and experiment for $A_{CP}$. 
Thus we strongly motivate some further experimental studies of $\Delta \Gamma_d$.

\section{Acknowledgements}
I would like to thank to the organizers for an interesting and motivating workshop. 
Many thanks to U. Nierste and G. Borissov for useful discussions concerning
the like-sign dimuon asymmetry. I am also grateful with A. Lenz for proofreading this manuscript
and with CONACyT, Mexico, for financially supporting my PhD program.


\begin{thebibliography}{99}
\bibitem{D01}
V. M. Abazov \textit{et al}. [D0 Collaboration], Phys. Rev. D \textbf{82}, 032001 (2010) [arXiv:1005.2757 [hep-ex]].
\bibitem{D02}
V. M. Abazov \textit{et al}. [D0 Collaboration], Phys. Rev. Lett. \textbf{105}, 081801 (2010) [arXiv:1007.0395 [hep-ex]].
\bibitem{D03}
V. M. Abazov \textit{et al}. [D0 Collaboration], Phys. Rev. D \textbf{84}, 052007 (2011) [arXiv:1106.6308 [hep-ex]].
\bibitem{D0_exp}
V. M. Abazov \textit{et al.}[D0 Collaboration], Phys. Rev. D \textbf{89}, 012002 (2014) [arXiv:1310.0447 [hep-ex]].
\bibitem{theo_Lenz_Uli}
A. Lenz and U. Nierste [arXiv:1102.4274 [hep-ph]].
\bibitem{theo_1}
A. Lenz and U. Nierste, JHEP \textbf{0706}, 072 (2007) [arXiv:0612167 [hep-ph]].
\bibitem{theo_2}
M. Beneke, G. Buchalla, A. Lenz, U. Nierste, Phys. Lett. B \textbf{576}, 173 (2003) [arXiv:0307344 [hep-ph]].
\bibitem{theo_3}
M. Ciuchini, E. Franco, V. Lubicz, F. Mescia, C. Tarantino, JHEP \textbf{0308}, 031 (2003) [arXiv:0308029 [hep-ph]].
\bibitem{theo_4}
M. Beneke, G. Buchalla, C. Greub, A. Lenz, U. Nierste, Phys. Lett. B \textbf{459}, 631 (1999) [arXiv:9808385 [hep-ph]]
\bibitem{Borissov}
G. Borissov and B. Hoeneisen, Phys. Rev. D \textbf{87}, 074020 (2013) [arXiv:1303.0175 [hep-ex]].
\bibitem{Delta_Gamma_d}
C. Bobeth, U. Haisch, A. Lenz, B. Pecjak and G. Tetlalmatzi-Xolocotzi, JHEP
\textbf{1406}, 040 (2014) [aXiv:1404.2531 [hep-ph]].
\bibitem{HFAG}
Y. Amhis \textit{et al.} [Heavy Flavour Averaging Group] [arXiv:1207.1158 [hep-ex]], updated results available at http://www.slac.stanford.edu/xorg/hfag/index.html. 
\bibitem{gamma}
J. Brod, A. Lenz, G. Tetlalmatzi-Xolocotzi and M. Wiebusch [arXiv:1412.1446 [hep-ph]].
\bibitem{Bobeth_Haisch}
C. Bobeth and U. Haisch, Acta Phys. Polon. B \textbf{44}, 127 (2013) [arXiv:1109.1826 [hep-ph]].
\bibitem{BABAR_bound}
B. Aubert \textit{et al.} [BaBar Collaboration], Phys. Rev. Lett. \textbf{96}, 241802 (2006) [arXiv:0511015 [hep-ex]].
\bibitem{Uli_dimuon}
U. Nierste, CKM 2014, https://indico.cern.ch/event/253826/contributions.
\bibitem{dimuon}
A. Lenz, G. Tetlalmatzi-Xolocotzi and B. Pecjak, to appear.
\end{thebibliography}
\end{document}